\documentclass[pra,aps,twocolumn,notitlepage,nofootinbib,longbibliography,superscriptaddress]{revtex4-1}%twocolumn,aps]
\usepackage{graphics,dcolumn}
\usepackage{bm}
\usepackage[fleqn]{amsmath}
\usepackage{hyperref}
\hypersetup{
     colorlinks=true,       % false: boxed links; true: colored links
    linkcolor=blue,          % color of internal links (change box color with linkbordercolor)
    citecolor=blue,        % color of links to bibliography
    filecolor=magenta,      % color of file links
    urlcolor=cyan           % color of external links
}
\usepackage{exscale,relsize}
\usepackage{epsfig}
\usepackage{amssymb}
\usepackage{amsmath}
\usepackage{euscript}
\usepackage{float}
\usepackage{tikz}
\usetikzlibrary{snakes}
\usetikzlibrary{decorations.pathmorphing}
\usetikzlibrary{patterns}
\usepackage{bbm}
\usepackage{footmisc}
\usepackage{epsfig}
\usepackage{amssymb}
\usepackage{amsmath}
\usepackage{euscript}
\usepackage{float}
 \usepackage{amsthm}

\usepackage[xcolor]{changes}
\definechangesauthor[color=orange]{VS}
\definechangesauthor[color=red]{ABU}
\definechangesauthor[color=purple]{DB}

\newcommand{\be}{\begin{equation}}
\newcommand{\e}{\end{equation}}

\newcommand{\beml}{\begin{subequations}}
\newcommand{\eml}{\end{subequations}}
\newcommand{\beq}{\begin{eqnarray}}
\newcommand{\eq}{\end{eqnarray}}
\newcommand{\ba}{\begin{array}}
\newcommand{\ea}{\end{array}}

\newcommand{\lt}{\left}
\newcommand{\rt}{\right}
\newcommand{\n}{\nonumber}

\newcommand{\la}{\langle}
\newcommand{\ra}{\rangle}

\newcommand{\re}{\,{\rm Re}\,}

\newcommand{\dif}{\mathrm{d}}
\newcommand{\pvec}[1]{\vec{#1}\mkern2mu\vphantom{#1}}

\begin{document}
\date{\today}

\title[Universal entanglement decay of photonic orbital angular momentum qubit states in atmospheric turbulence: an analytical treatment]{Universal entanglement decay of photonic orbital angular momentum qubit states in atmospheric turbulence: an analytical treatment}

\author{David Bachmann, Vyacheslav N. Shatokhin and Andreas Buchleitner}
\address{Physikalisches Institut, Albert-Ludwigs-Universit\"at Freiburg, Hermann-Herder-Str. 3,
D-79104 Freiburg, Federal Republic of Germany}

\begin{abstract}
We study the entanglement evolution of photonic orbital angular momentum qubit states with opposite azimuthal indices $l_0$, in a weakly turbulent atmosphere. Using asymptotic methods, we deduce analytical expressions for the amplitude of turbulence-induced crosstalk  between the modes $l_0$ and $-l_0$. Furthermore, we analytically establish distinct, universal entanglement decay laws for Kolmogorov's turbulence model and for two approximations thereof. 
\end{abstract}

%\pacs{
%03.67.Hk, 
%42.50.Tx,
%42.68.Bz
%}
\maketitle

%\tableofcontents

\section{Introduction}
\label{sec:intro}
Helical wave fronts of twisted photons allow for encoding high-dimensional (qudit) states in the orbital angular momentum (OAM) degree of freedom that is characterized by an azimuthal index $l_0=0,\pm 1,\pm 2, \ldots$ \cite{PhysRevA.45.8185,Molina-Terriza:2007zr,franke-arnold08}.  Using qudit states  can potentially ensure higher channel capacity \cite{Gibson:04,Barreiro:2008cr} and enhanced security \cite{PhysRevLett.85.3313,PhysRevA.64.012306} of quantum communication channels as compared to two-dimensional, polarization-based, encoding. So far, the full-scale use of twisted photons in free space quantum communication has remained elusive due to the sensitivity of the photons' wave fronts with respect to intrinsic fluctuations of the refractive index of air (atmospheric turbulence) \cite{tatarskij}. Notwithstanding, recent experimental progress where OAM encoding was used in free space quantum key distribution over up to 300 m \cite{vallone14,Sit:17}, for entanglement distribution over 3 km \cite{Krenn17112015}, and for the transmission of classical 
twisted light over 143 km \cite{Krenn29112016}, indicates that turbulence-induced phase-front distortions do not in principle preclude the reliable transfer of twisted photons through the atmosphere. Yet, we still lack a complete understanding of the behaviour of photonic OAM states in turbulence, even for OAM-encoded qubits.% and inspires further progress of this research area. 

Since one of the most prominent quantum communication protocols \cite{PhysRevLett.67.661} is based on quantum entanglement \cite{Mintert05b,horodecki09} -- which is readily generated between OAM states in the lab \cite{Mair:2001kx,Fickler02112012,woerdman12} -- experimental \cite{Pors:11,alpha13,roux16,goyal2016,Ndagano:2017mz} and theoretical \cite{gopaul07,Sheng:12,roux11,PhysRevA.90.052115,roux15,PhysRevA.95.023809} efforts explore how entanglement of twisted biphotons decoheres in the atmosphere. In particular, it was predicted \cite{raymer06} -- and recently confirmed experimentally \cite{alpha13} -- that entangled qubit states with opposite OAM become more robust in weak turbulence as the azimuthal index $l_0$ increases.  

Some useful insight into this feature was provided through the introduction of the {\it phase correlation length} $\xi(l_0)$ \cite{leonhard15} that is the characteristic length scale associated with the transverse spatial structure of OAM beams. In particular, %the entanglement decay of photonic OAM qubit states in a weakly turbulent atmosphere modeled by a single phase screen \cite{BUCKLEY19751431} is {\it universal} and governed solely by the effective turbulence strength $\xi(l_0)/r_0$ \cite{leonhard15}, where $r_0$ is the turbulence correlation length \cite{FRIED:66}. Although 
it was numerically shown \cite{leonhard15} that the individual temporal evolution of the concurrence \cite{Mintert05b,wootters98} of initially maximally entangled OAM qubit states with quantum numbers $\pm l_0$ collapses onto an $l_0$-independent, universal entanglement decay, upon rescaling the turbulence strength by $\xi(l_0)$. However, no analytical derivation of this universal decay law has so far been available.

%entanglement evolution (as characterized by concurrence \cite{Mintert05b,wootters98}) of entangled states with distinct values of $l_0$ collapse onto one universal entanglement decay curve, upon rescaling turbulence strength with $\xi(l_0)$.   

%An analytical formula linking entanglement to the relative crosstalk amplitude $\tilde{b}$ (see Sec. \ref{sec:gen}) characterizing  coupling between modes with indices $-l_0$ and $l_0$ was found in \cite{leonhard15}. Although this result clearly shows that an increase of $\tilde{b}$ accelarates entanglement decay, the explicit form of $\tilde{b}$ as a function of $\xi(l_0)/r_0$ was not derived. 

In the present contribution, we revisit the entanglement evolution of photonic OAM qubit states in weak turbulence and generalize the results of \cite{leonhard15} along two directions: First, we deduce the entanglement decay law analytically. To this end, we exploit the universality of the decay, and consider the limit $l_0\gg 1$, using asymptotic methods.  Second, we 
derive explicit expessions for the universal decay law for 
three distinct models of turbulence -- distinguished by the exponent $\alpha$ of the phase structure function \cite{tatarskij,andrews} (see Sec.~\ref{sec:model}). This generalization is inspired by the fact that the universal form of the entanglement decay relies on the weakness of the turbulence. The latter assumption implies that the impact of turbulence on the propagated wave can be described by a single phase screen \cite{BUCKLEY19751431}, irrespective of the specific turbulence model. Since variation of $\alpha$ leads to different types of atmospheric disorder, it is interesting to explore, somewhat in the spirit of \cite{PhysRevX.6.031023}, how this affects entanglement evolution in turbulence. %Furthermore, consideration of different values of $\alpha$ has practical applications in the upper troposphere and stratosphere, where non-Kolmogorov models of turbulence play an important role \cite{toselli2008}.   

The paper is structured as follows: In the next section we present our general model of turbulence, parametrized by the exponent $\alpha$, as well as our method to access the entanglement evolution. In Sec. \ref{sec:universal} we derive explicit expressions for the relative crosstalk amplitude, $\tilde{b}$, in the asymptotic limit $l_0\gg 1$, for $\alpha=1,5/3,2$, and compare these asymptotic results to numerically exact data. We finally give analytic formulations for the universal entanglement decay law, for those different values of $\alpha$. Section \ref{sec:concl} concludes the manuscript.  

\section{Model}
\label{sec:model}
We consider a maximally entangled OAM qubit state of two photons, encoded in two Laguerre-Gauss (LG) modes with radial and azimuthal quantum numbers $p=0$ and \mbox{$l=\pm l_0$}, respectively, and relative phase $\varphi$: 
\begin{equation}
|\psi_0\ra=\frac{1}{\sqrt{2}}(|l_0\ra|-l_0\ra+e^{i\varphi}|-l_0\ra|l_0\ra),
\label{in_state}
\end{equation}  
Such states can be generated experimentally  \cite{Mair:2001kx,woerdman12}, and large values up to $l_0=300$ were reported \cite{Fickler02112012}. The entangled photons are sent, in opposite directions along the (horizontal) $z$-axis, through the atmosphere and the impact of the latter on the transmitted photons is modelled by two independent phase screens --- one for each photon. This model of atmospheric turbulence ignores turbulence-induced intensity fluctuations. Furthermore, we neglect diffraction, assuming that the widths of the LG beams remain constant along the propagation paths. As follows from \cite{paterson04}, both assumptions are valid for short propagation distances of about one km for each photon. Note that this is a standard approach \cite{andrews} which only accounts for scattering of an optical wave on refractive index inhomogeneities of a typical scale much larger than the optical wave length. This model does therefore not describe any loss of information as induced, e.g., by photon absorption, Rayleigh scattering by air molecules, or Mie scattering by aerosols. These processes, however, only affect the total transmission rate of the photons, but leave the description of the atmospheric imprint on the successfully transmitted photons invariant.
%On the other hand, photonic OAM is conserved under forward scattering  \cite{andrews11} and, therefore, the latter cannot trigger the additional coupling, or crosstalk, among the OAM modes, on top of addressed below crosstalk that is induced by phase aberrations.}

When a pair of twisted photons prepared in state (\ref{in_state}) is sent across the atmosphere, turbulence-induced random phase shifts lead to the coupling, or crosstalk, of the initially excited (LG$_{0\pm l_0}$) modes to other LG$_{pl}$ modes, spreading, in general, over infinitely many values of $p$ and of $l$. Upon averaging this high-dimensional pure state over independent statistical realizations of those random shifts, one ends up with a mixed state. 

Thus, the output biphoton state $\varrho$ can be described by a map \cite{leonhard15}
\begin{equation}
\varrho=(\Lambda_1\otimes\Lambda_2)\varrho_0,
\label{map}
\end{equation} 
where $\Lambda_i$ is the disorder-averaged turbulence map acting on the state of the $i$th photon, and $\varrho_0=|\psi_0\ra\la\psi_0|$. The tensor product expresses the independence of the action of the two screens on either photon.

Since we are only interested in the OAM entanglement evolution in the subspace of the initially populated OAM modes, we trace over the radial quantum number $p$, and project the output state onto the subspace spanned by the four product vectors:
$|l_0\ra|l_0\ra,|l_0\ra|-l_0\ra,|-l_0\ra|l_0\ra,|-l_0\ra|-l_0\ra$ \cite{leonhard15}.
Due to the projection, the norm of the resulting state is reduced and the state needs to be renormalized by its trace \cite{hiesmayr03}. The entanglement evolution of the thereby obtained mixed, bipartite qubit state can be evaluated using Wootter's concurrence \cite{wootters98}.

The elements of the map $\Lambda_i$ required for the evaluation of the output state $\varrho$ are given by (see \ref{Appmap})
\begin{align}
\Lambda_{l,\pm l}^{0,l_0,0,l_0'}&=\frac{1}{2\pi}\delta_{l_0-l_0',l\pm l}\int_{0}^{\infty}\hspace{-8pt}\int_{0}^{2\pi} r \dif r \dif \vartheta\n\\
&\times R_{0,l_0}(r) R_{0,l_0'}(r)
	e^{-i\frac{\vartheta}{2}[(l\pm l)-(l_0+l_0')]}\n\\
	&\exp\left\{-\frac{1}{2}D_\phi\left[2r\sin\left(\frac{\vartheta}{2}\right)\right]\right\},
	\label{defLambda}
	\end{align}
where upper and lower indices of $\Lambda_{l,\pm l}^{0,l_0,0,l_0'}$ correspond to the input and output states, respectively, $\delta_{l_0-l_0',l\pm l}$ is the Kronecker delta,
\begin{equation}
R_{0,l_0}(r)=\frac{2}{\sqrt{|l_0|!}}\frac{1}{w_0}\left(\frac{\sqrt{2}r}{w_0}\right)^{|l_0|}\,e^{-r^2/w_0^2},
\label{eq:radial}
\end{equation}
is the radial part of the LG$_{0\pm l_0}$ mode at $z=0$, with $p=0$, $l=\pm l_0$, and $w_0$ is the mode's width. Furthermore,
\begin{equation}
D_\phi(x)=\gamma\left(\frac{x}{r_0}\right)^\alpha
\label{eq:pls}
\end{equation}
is the phase structure function of turbulence, which determines the statistics of the spatial wavefront deformations \cite{FRIED:65}. Here, $\gamma=6.88$ and
\begin{equation}
r_0=(0.423\,C_n^2k^2L)^{-3/5}
\label{eq:fried}
\end{equation}
is the Fried parameter \cite{FRIED:66,andrews}, with $C_n^2$ the index-of-refraction structure constant, $k$ the optical wave number, and $L$ the propagation distance. Throughout this work, we assume that both photons traverse equal-length paths in the atmosphere characterized by the same $C_n^2$. Hence, $r_0$ coincides for both screens.

 The exponent $\alpha$ in Eq. \eqref{eq:pls} specifies the turbulence model. In particular, $\alpha=5/3$ characterizes Kolmogorov's description  \cite{tatarskij,andrews}, but we subsequently also consider the entanglement evolution for $\alpha=1,2$ which can be regarded as integer-value ``approximations'' of $\alpha=5/3$. 
In fact, the phase structure function with $\alpha=2$ is oftentimes used in the literature \cite{raymer06,PhysRevA.95.023809,roux15,Ndagano:2017mz}, with Eq. \eqref{eq:pls} then called the ``quadratic
approximation'' \cite{Leader:78} of the Kolmogorov phase structure function (by analogy, we refer to the case $\alpha=1$ as the ``linear approximation''). Apart from the fact that the quadratic approximation allows for obtaining analytical results for entanglement decay of OAM qubit states even under conditions of strong turbulence \cite{roux15,PhysRevA.95.023809}, it is sometimes considered to be slightly pessimistic \cite{6695762} as compared to the Kolmogorov model. Indeed, as follows from Eqs. (\ref{defLambda}) and (\ref{eq:pls}), the matrix elements of the map $\Lambda_i$ governing the evolution of the density operator will decay faster with $x$ for $\alpha=2$ than for $\alpha=5/3$. Consequently, the first-order spatial correlation function of the propagated field (also known as the mutual coherence function \cite{tatarskij,andrews}) decreases more rapidly within the square-law approximation, which is considered \cite{6695762} to be harmful for optical communication. However, as for the entanglement evolution of OAM states, a smaller magnitude of those matrix elements of $\Lambda_i$ that reflect the coupling between OAM modes with $-l_0$ and $l_0$ is actually useful, because it is the intermodal crosstalk that causes entanglement decay \cite{leonhard15}.

As a result, as we show in this work, the robustness of entanglement is most pronounced for $\alpha=2$. From this viewpoint, the quadratic approximation appears as overly optimistic. Consistently, the linear approximation, $\alpha=1$, yields pessimistic results as compared to those derived from the Kolmogorov model.

\section{Universal entanglement decay laws}
\label{sec:universal}
\subsection{General expressions}
\label{sec:gen}

Upon application of the turbulence map \eqref{map}, the output state's concurrence is given analytically by \cite{leonhard15}
\begin{equation}
C(\varrho)=\max\lt[0,\frac{1-2\tilde{b}}{(1+\tilde{b})^2}\rt], 
\label{conc}
\end{equation}
where $\tilde{b}=b/a$ is the reduced, or relative, crosstalk amplitude, with
\begin{subequations}
\begin{align}
a=\Lambda^{l_0,l_0}_{l_0,l_0}&=\Lambda^{-l_0,-l_0}_{-l_0,-l_0}=\Lambda^{-l_0,l_0}_{-l_0,l_0}=\Lambda^{l_0,-l_0}_{l_0,-l_0},\label{ind_a}\\
b&=\Lambda^{-l_0,-l_0}_{l_0,l_0}=\Lambda^{l_0,l_0}_{-l_0,-l_0},\label{ind_b}
\end{align}
\end{subequations}
expressed in terms of the phase screen map's matrix elements \eqref{defLambda} (we drop their upper indices $p=0$ for brevity). We note that Eq. \eqref{conc} is valid for a single phase screen description of turbulence, whereas the specific turbulence model is included via the dependence of the parameters $a$ and $b$ on $\alpha$. As follows from the definition \eqref{defLambda}, in the absence of turbulence, or, equivalently, when $r_0\rightarrow \infty$, the survival amplitude tends to one, $a\rightarrow 1$,  while $b\rightarrow 0$, for the crosstalk amplitude. Hence, also $\tilde{b}\rightarrow 0$, and concurrence converges to its maximum value of unity. In contrast, when turbulence comes into play, crosstalk between different OAM modes is expected, and $C(\varrho)$ given by \eqref{conc} should thus drop below unity. As we show below, this expectation is not always justified.

In the framework of the Kolmogorov model of turbulence, by numerical assessment  of $a$ and $b$, we established earlier that the behaviour of $C(\varrho)$ is {\it universal} in the sense that the a priori $l_0$-dependent entanglement evolution with the penetration depth into the turbulent medium collapses onto one universal curve, for arbitrary OAM $l_0>1$, provided $C(\varrho)$ is considered as a function of the rescaled turbulence strength $\xi(l_0)/r_0$ \cite{leonhard15}. Here, $\xi(l_0)$ is the {\it phase correlation length} of an LG$_{0l_0}$ beam of width $w_0$, which reads \cite{leonhard15}
\begin{equation}
\xi(l_0)=\sin\lt(\frac{\pi}{2|l_0|}\rt)\frac{w_0}{\sqrt{2}}\frac{\Gamma(|l_0|+3/2)}{\Gamma(|l_0|+1)}.
\label{corr_length}
\end{equation}
Thus, the entanglement evolution within the Kolmogorov and single phase screen model is determined by the ratio of the only two characteristic length scales, of which $\xi(l_0)$ refers to the OAM beam and $r_0$ to the turbulent medium. In the following, we analytically show that the universality of the entanglement decay persists also for $\alpha=1$ and $\alpha=2$, though with a different functional form, which we identify for $\alpha=1,5/3,2$.

By virtue of Eq.~(\ref{conc}), the universality of the concurrence decay law originates from that of the relative crosstalk amplitude, $\tilde{b}\equiv\tilde{b}(\xi(l_0)/r_0)$. Therefore, we need to infer the explicit form of the latter. To that end, we consider the limit $l_0\gg 1$, and apply asymptotic analysis. Using the method of steepest descent \cite{carrier2005functions}, we obtain the following expressions (see \ref{appA}):
\begin{align}
a&\approx\frac{1}{\pi}A^{-1/\alpha}\,\Gamma\lt(1+\frac{1}{\alpha}\rt),\label{asa}\\
b&\approx\frac{1}{\pi}\re\lt[\int_0^\infty \dif{x}\exp(-A\,x^\alpha-2il_0\,x)\rt]\label{asb},
\end{align}
where $A= 2^{-\alpha-1}\,\gamma\,(2l_0)^{\alpha/2}\,t^\alpha$, and 
\begin{equation}
t:= w_0/r_0
\end{equation}
is the turbulence strength. The asymptotic solutions (\ref{asa}) and (\ref{asb}) essentially accomplish our target: Indeed, \eqref{asa} already has an explicit analytical form. Equation \eqref{asb} involves a highly oscillatory integral which, in turn, can be evaluated by asymptotic methods. Though, for any $\alpha=l/k$, with $l,k$ positive integers, Eq. \eqref{asb} can be evaluated exactly and is given in terms of either elementary ($\alpha=1,2$) or of special ($\alpha=5/3$) functions. Next, we analyze these three cases in more detail.

Before proceeding, we would like to point out that any distinctions in the $l_0$- and $t$-dependencies of $\tilde{b}$ and, hence, of the entanglement decay laws, for different $\alpha$, can stem exclusively from the turbulence-induced crosstalk amplitude $b$. Indeed, as follows from our asymptotic expression \eqref{asa}, for {\it any} $1\leq \alpha\leq 2$, the survival amplitude $a\sim 1/t\sqrt{l_0}$. Finally, to study the behaviour of the survival amplitude for $0<t\ll 1$, using the approximate formula \eqref{asa}, the physical bounds $0< a\leq 1$ have to be kept in mind, and  $l_0\gg 1$ needs to be chosen accordingly.

\subsection{Asymptotic expressions for survival and crosstalk amplitudes}
\subsubsection{Linear approximation ($\alpha=1$)}
\label{sec:lin}
For $\alpha=1$, \eqref{asa} and \eqref{asb} yield the following asymptotic results ($l_0\gg 1$):
\begin{equation}
a\approx\frac{4}{\pi\gamma}\frac{1}{\sqrt{2l_0}t}, \quad b\approx \frac{\gamma}{4\pi}\frac{\sqrt{2l_0}t}{4l_0^2+\gamma^2 t^2l_0/8}.
\label{ablin}
\end{equation}
%where $t\equiv w_0/r_0$ is the dimensionless turbulence strength. 
%\begin{widetext}
\begin{center}
\begin{figure}
{\includegraphics[width=7cm]{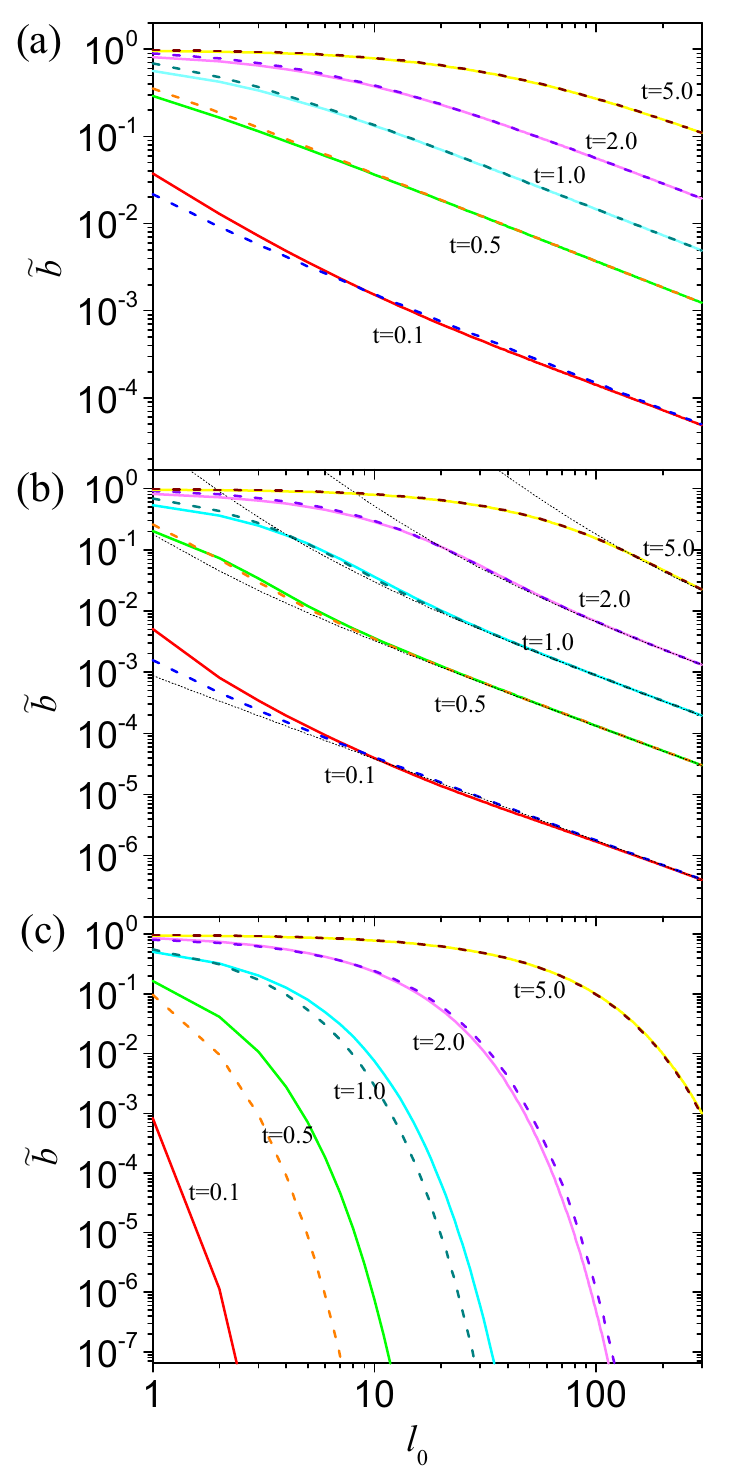}}
\caption{(Color online) Relative crosstalk amplitude $\tilde{b}$ versus the azimuthal index $l_0$ (on a log-log scale), for $t\equiv w_0/r_0=0.1,0.5,1.0,2.0,5.0$ and for three values of $\alpha$ [see Eq. (\ref{eq:pls})]: (a) $\alpha=1$, (b) $\alpha=5/3$, \mbox{(c) $\alpha=2$}. Solid lines indicate exact numerical integration of Eq. (\ref{defLambda}), dashed lines correspond to asymptotic expressions (a) (\ref{ablin}), (b) (\ref{abK}), (c) (\ref{as_a_b_q}). Thin dotted lines in panel (b) represent the asymptotic expansion (\ref{As_ser1}). }
\label{fig1}
\end{figure}
\end{center}
In Fig.~\ref{fig1}(a) we show the function $\tilde{b}=b/a$, as obtained by exact numerical evaluation of the integrals in Eq. (\ref{defLambda}), together with the asymptotic solution derived from Eq. (\ref{ablin}). We notice that the asymptotic value of the ratio $b/a$ provides excellent agreement with the exact numerical solution already for $l_0\gtrsim 10$. As follows from Eq. (\ref{ablin}), at leading order the relative crosstalk amplitude reads
\begin{equation}
\tilde{b}\sim \frac{\gamma^2t^2}{32 l_0},
\label{tildeb_lin}
\end{equation}
i.e., is proportional to $t^2/l_0$.

\subsubsection{Kolmogorov turbulence ($\alpha=5/3$)}
\label{kolm_l0}
After a suitable coordinate transformation  (see \ref{appA}), the integral in Eq. (\ref{asb}) reduces to a tabulated integral \cite{Prudnikov:1992:IS4}. 
Explicitly, the amplitudes $a$ and $b$ for $\alpha=5/3$ read:
\begin{subequations}
\begin{align}
a&\approx\frac{\sqrt{2}(\gamma/2)^{-3/5}}{\pi t\sqrt{l_0}}\Gamma\lt(\frac{8}{5}\rt), \label{aK}\\
b&\approx \frac{\sqrt{15}}{2^3\pi^4}\label{bK}\\
&\times\re\lt[\frac{e^{-i\pi/5}}{s}G^{3,5}_{5,3}\lt(\lt.\frac{5^5q^3}{3^3s^5}\rt|\begin{array}{lllll}
0,&\frac{1}{5},&\frac{2}{5},&\frac{3}{5},&\frac{4}{5}\\
0,&\frac{1}{3},&\frac{2}{3}&&
\end{array}
\rt)\rt],\n
\end{align}
\label{abK}%
\end{subequations}
where $s=2il_0\,e^{-i\pi/5}$, $q=\gamma\, 2^{-11/6}\,t^{5/3}\,l_0^{5/6} e^{-i\pi/3}$, and $G^{3,5}_{5,3}(x|\dots)$ is the Meijer G-function, which can be expressed through  generalized hypergeometric functions \cite{bateman}. As in the case $\alpha=1$, the numerically exact and the asymptotic results merge for $l_0\gtrsim 10$ (see Fig. ~\ref{fig1}(b)). However, the hypergeometric functions are a representation of an infinite series \cite{abramowitz1966handbook}. A transparent analytical form of the dependence of $\tilde{b}$ on $l_0$ can be deduced from the {\it asymptotic series} expansion of $\tilde{b}$ which can be derived directly from (\ref{asa},\ref{asb}). For a {\it large} value of a parameter (in our case, for $l_0\gg 1$) the subsequent terms of an asymptotic series first decrease in magnitude, but then start progressively increasing \cite{dingle}.\footnote{This contrasts the behaviour of convergent series representing an analytic function $f(z)$ around its regular point $z=z_0$.} Thus, the asymptotic series are divergent. Nonetheless, these series are useful, since a {\it finite} number of terms of an asymptotic series expansion yields a very accurate representation of the function at large parameter values \cite{dingle}. As regards the relative crosstalk amplitude $\tilde{b}$, its best approximation for $ 0.1\lesssim t\lesssim 5$ and $l_0\gtrsim 150$ is obtained by the first three terms of the asymptotic series (see Eqs. \eqref{tildebK}-\eqref{As_ser} in \ref{appA}):
\begin{align}
\tilde{b}&\approx 0.380\!\lt(\!\frac{t^2}{l_0}\!\rt)^{\!\!\frac{4}{3}}\!\!\!\!+\!1.231\!\lt(\!\frac{t^2}{l_0}\!\rt)^{\!\!\frac{13}{6}}\!\!\!\!\!\!+\!3.735\!\lt(\!\frac{t^2}{l_0}\!\rt)^{\!\!3}\!\!\!+\!O\!\!\lt[\lt(\!\frac{t^2}{l_0}\!\rt)^{\!\frac{23}{6}}\rt].
\label{As_ser1}
\end{align} 
It is noteworthy that, as in the linear-approximation-scenario (see Sec. \ref{sec:lin}), the relative crosstalk amplitude is represented as a power series in $t^2/l_0$. At leading order, $\tilde{b}$ exhibits a power-law behaviour $\sim (t^2/l_0)^{4/3}$, which is faster than the linear dependence (\ref{tildeb_lin}) characteristic of the linear approximation. For instance, the values of the relative crosstalk, at $l_0=150$ and at the minimum turbulence strength $t=0.1$ presented in Fig. \ref{fig1}, are $\approx 10^{-4}$ and $\approx 10^{-6}$ for $\alpha=1$ and $\alpha=5/3$, respectively. 
%Nonetheless, even for the smallest turbulence strength $t=0.1$ presented in Fig.~\ref{fig1}, the relative crosstalk amplitude remains finite ($\tilde{b}\simeq 10^{-6}$) for $l_0=100$, according to the convention formulated in Sec. \ref{sec:lin}. 

\subsubsection{Quadratic approximation ($\alpha=2$)}
\label{sec_b_quad}
In contrast to the above cases $\alpha=1,5/3$, for $\alpha=2$ the relevant elements of the map $\Lambda_i$, defined by Eq. (\ref{defLambda})-(\ref{eq:pls}), can be found in tables of integrals \cite{prudnikov1990integrals}. Setting the upper and lower indices in Eq. \eqref{defLambda} according to Eqs. (\ref{ind_a}) and (\ref{ind_b}), we obtain the following exact results for the amplitudes $a$ and $b$ in (\ref{asa},\ref{asb}), respectively: 
\begin{subequations}
\begin{align}
a&=2^{l_0+1}(2+\tau)^{-l_0-1}\n\\
&\times\!F\lt(\frac{2l_0+1}{2},\frac{2l_0+2}{2};1; \lt(\frac{\tau}{2+\tau}\rt)^2\rt),\label{exact_a_q}\\
b&=2^{l_0+1}(2+\tau)^{-3l_0-1}\lt(\frac{\tau}{2}\rt)^{2l_0}\lt(
\begin{array}{c}
3l_0\\
l_0
\end{array}
\rt)\n\\
&\times\!
F\lt(\frac{3l_0+1}{2},\frac{3l_0+2}{2};2l_0+1;\lt(\frac{\tau}{2+\tau}\rt)^2\rt),\label{exact_b_q}
\end{align}
\label{quad}%
\end{subequations}
where $F(\beta,\delta;\eta;z)$ is the hypergeometric function \cite{abramowitz1966handbook} and $\tau\equiv \gamma t^2$. 

Since solutions in terms of hypergeometric functions provide poor insight, we inspect the approximate forms of Eqs. (\ref{exact_a_q}, \ref{exact_b_q}), using the asymptotic formulas (\ref{asa}), (\ref{asb}). 
For $\alpha=2$, these produce the compact expressions
\begin{equation}
a\approx \frac{1}{t\sqrt{\pi \gamma l_0}}, \quad b\approx \frac{1}{t\sqrt{\pi \gamma l_0}}\exp\lt[-\frac{4l_0}{\gamma t^2}\rt].
\label{as_a_b_q}
\end{equation}
We remark that, according to Eq. (\ref{as_a_b_q}) and Fig.~\ref{fig1}(c), both, the asymptotic and the exact solutions for $\tilde{b}$ exponentially decay versus $l_0$; furthermore, they 
tend to coalesce for $l_0\gtrsim 150$ and $t> 2.0$. For $t\leq 2.0$, the decay of the asymptotic $\tilde{b}$ versus $l_0/t^2$ in Eq. (\ref{as_a_b_q}) is faster than the exact one given by \eqref{quad}. Note, however, that the deviations occur at exact and approximate values which are fairly small. For instance, at $l_0=150$, the approximate and the exact results at $t=2.0$ are $\tilde{b}\approx 3.4\times 10^{-10}$ and $1.3\times 10^{-9}$, respectively, and can be neglected. At $t<2.0$, the exact and the approximate values of $\tilde{b}$ become even tinier. 

The exponential dependence of the relative crosstalk amplitude on the ratio $l_0/t^2$, see Eq. (\ref{as_a_b_q}), indicates a qualitative difference between the present and the previous, $\alpha=1,5/3$, cases, both characterized by slow, power-law dependencies of $\tilde{b}$ on $t^2/l_0$. Indeed, owing to this slow decay for $\alpha=1,5/3$, a finite crosstalk amplitude between, widely separated, OAM modes $l_0$ and $-l_0$ is induced already for weak turbulence strength $t\simeq 0.1$. In contrast, for $\alpha=2$ such a coupling is exponentially suppressed even for moderate $t\simeq 2.0$. 
As we will see in Sec.~\ref{sec:ent}, the consequence of this suppression is that photonic OAM-entangled qubit states are most robust under turbulence characterized by $\alpha=2$. 
%\added[id=VS]{Namely, concurrence plotted vs turbulence strength exhibits a plateau, which gets wider with increasing $l_0$, where $C(\varrho)\approx 1$. This feature can be found in, both, analytical calculations \cite{raymer06} and numerical simulations \cite{alpha13,Yan:16}.}
%For that purpose, in the next section we study the behavior of the relative crosstalk amplitude and concurrence as functions of the rescaled turbulence strength, $\xi(l_0)/r_0$.
%%%%%%%%%%%%%%%%%%%%%%%%%%%%
\begin{center}
\begin{figure}
{\includegraphics[width=7cm]{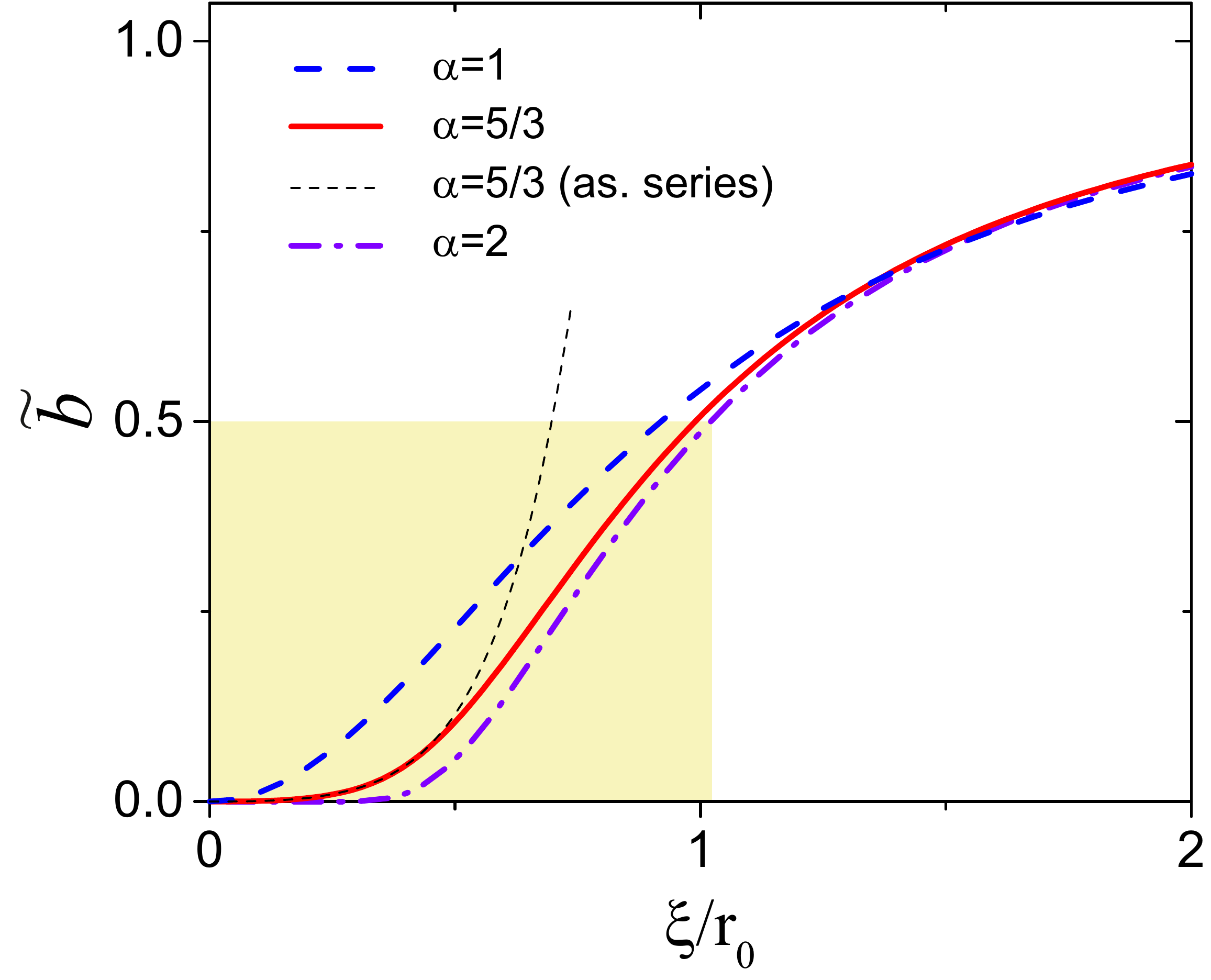}}
\caption{(Color online) Universal relative crosstalk amplitude $\tilde{b}$ versus $\xi/r_0$, for $\alpha=1$ [Eq. (\ref{tildeb1})], $\alpha=5/3$ [the universal function, Eqs. (\ref{tbK}), and the asymptotic series, Eq. (\ref{As_ser})], and $\alpha=2$ [Eq. (\ref{tbquad})]. The shaded area highlights the parameter range $\tilde{b}\leq 1/2$ and $\xi/r_0\lesssim 1$, where the concurrence in Fig.~\ref{fig3} remains finite, due to \mbox{Eq. (\ref{conc})}.}
\label{fig2}
\end{figure}
\end{center}
%%%%%%%%%%%%%%%%%%%%%%%%%%%%%

\subsection{Rescaled relative crosstalk amplitude and concurrence}
\label{sec:ent}

\begin{center}
\begin{figure}
{\includegraphics[width=7cm]{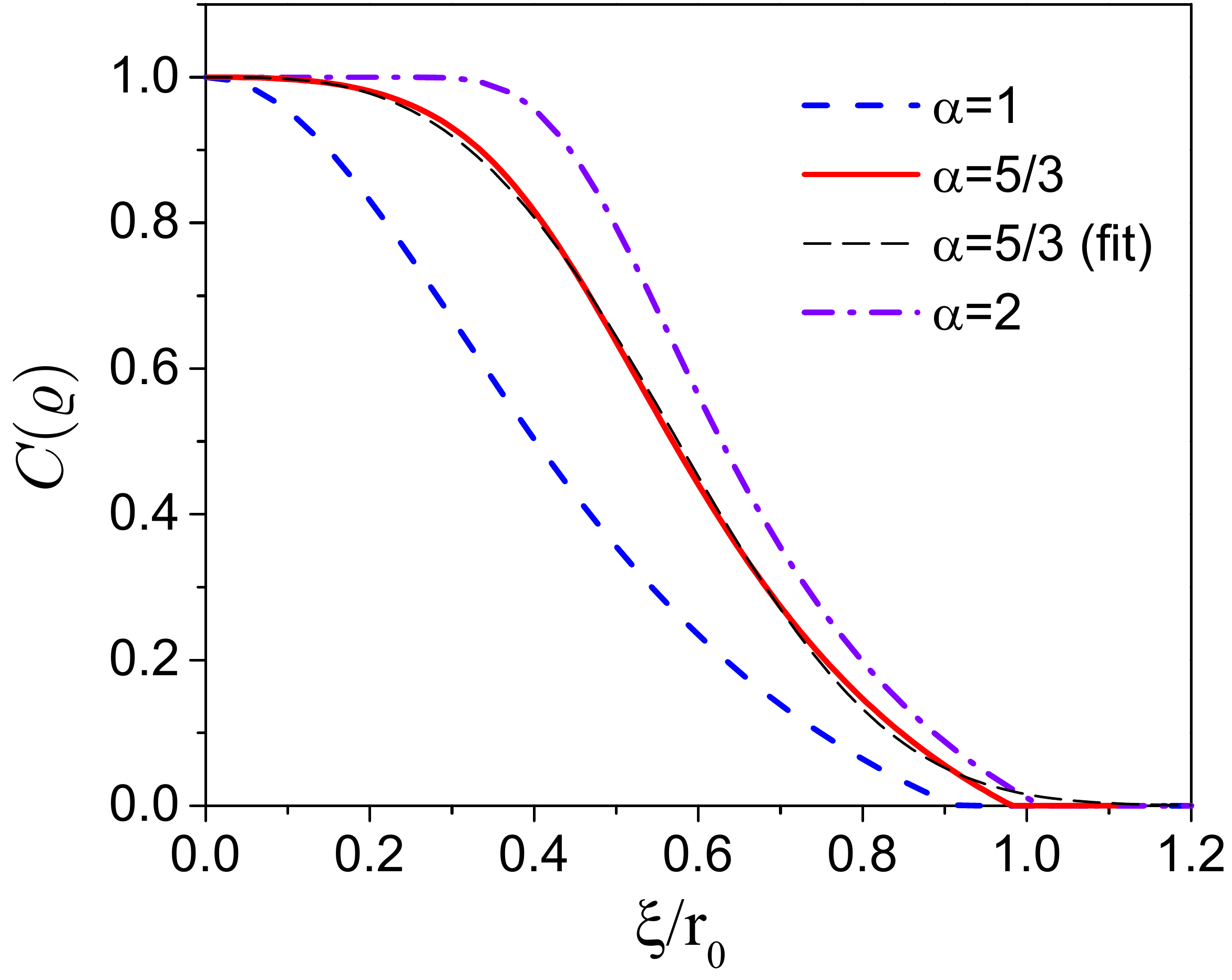}}
\caption{(Color online)  Universal decay of $C(\varrho)$ versus $\xi/r_0$, for $\alpha=1,5/3,2$, obtained from 
(\ref{conc}), with $\tilde{b}$ given, respectively, by Eqs. (\ref{tildeb1}), (\ref{tbK}) and (\ref{tbquad}). The thin dashed line that is nearly overlapping with the solid line represents the fitting function $g(\xi/r_0)=\exp[-4.16(\xi/r_0)^{3.24}]$ employed in \cite{leonhard15}.}
\label{fig3}
\end{figure}
\end{center}

In the limit $l_0\gg 1$, the expression for the phase correlation length given in Eq. \eqref{corr_length} can be simplified by recalling the asymptotics of the sine, $\sin\lt(\pi/2l_0\rt)\sim \pi/2l_0$,  
%\be
%\sin\lt(\frac{\pi}{2|l_0|}\rt)\sim \frac{\pi}{2|l_0|},
%\e
and of
the gamma-function \cite{abramowitz1966handbook}:
\begin{equation}
\Gamma(l_0+x)\sim \sqrt{2\pi}e^{-l_0}l_0^{l_0+x-1/2}.
\end{equation}
One obtains
\begin{equation}
\xi=\xi(l_0)\sim \frac{\pi}{2\sqrt{2}}\frac{w_0}{\sqrt{l_0}},
\label{asxi}
\end{equation}
which can be employed in Eqs. (\ref{ablin}, \ref{abK}, \ref{as_a_b_q}) to express the rescaled crosstalk in terms of $\xi/r_0$, yielding the sought-after universal behaviour.
%By using the above formula, Eqs. (\ref{ablin}), \eqref{abK}, and \eqref{as_a_b_q} turn into functions of the ratio $\xi/r_0$ (for brevity, from now on we skip the argument of $\xi$), thereby yielding the explicit shape of the universal curves. Let us present these universal functions.
We find for $\alpha=1$,
\begin{equation}
\tilde{b}=\frac{\lt(\gamma \xi/2r_0\rt)^2}{\pi^2+\lt(\gamma \xi/2r_0\rt)^2};
\label{tildeb1}
\end{equation}
for $\alpha=5/3$  (see \ref{appA}),
\begin{align}
\tilde{b}&=\sqrt{\frac{3}{5}}\frac{3^{3/5}}{(2\pi)^3}\frac{1}{\Gamma\lt(\frac{8}{5}\rt)}\label{tbK}\\
&\times\!\!\re\!\!\lt[
G^{3,5}_{5,3}\lt(\Biggl.\frac{-i\gamma^35^5\xi^5}{\pi^56^3r_0^5}\Biggr|\begin{array}{lllll}
\frac{1}{5},&\frac{2}{5},&\frac{3}{5},&\frac{4}{5},&1\\
\frac{1}{5},&\frac{8}{15},&\frac{13}{15}&&
\end{array}
\rt)
\rt]\nonumber\\
&\sim 0.29(\xi/r_0)^{8/3},
%\label{tbKfull}
\label{bKappr}
\end{align}
with the bottom line of \eqref{bKappr} valid in the limit $\xi/r_0\ll 1$, see Eqs. (\ref{As_ser1}) and \eqref{asxi},
%\be
%\tilde{b}\sim 0.29(\xi/r_0)^{8/3}.
%\label{bKappr}
%\e
and for $\alpha=2$,
\begin{equation}
\tilde{b}=\exp\lt(-\frac{\pi^2 r_0^2}{2\gamma \xi^2}\rt).
\label{tbquad}
\end{equation}
It is easy to check that in all three cases, Eqs. (\ref{tildeb1}\,--\,\ref{tbquad}), 
$\tilde{b}(0)=0$ and $\tilde{b}(\infty)=1$, where the latter value indicates the turbulence-induced {\it saturation} of the relative crosstalk amplitude. However, the behaviour of $\tilde{b}(\xi/r_0)$ is rather distinct in the range $0<\xi/r_0\lesssim 1$ (see Fig.~\ref{fig2}) for the three models of turbulence, what implies different forms of the entanglement decay (see Fig.~\ref{fig3}). 

Figure~\ref{fig2} shows that the rescaled relative crosstalk amplitude $\tilde{b}$ is maximal for $\alpha=1$, and minimal for $\alpha=2$, in the range  $0<\xi/r_0\lesssim 1$. Moreover, for $\alpha=2$, the amplitude $\tilde{b}$ remains vanishingly small for $\xi/r_0\lesssim 0.3$. This is a consequence of the exponential suppression of the crosstalk in the neighbourhood of $\xi/r_0=0$ [see Eq.~(\ref{tbquad})]. 
% and effectively behaves as the function most slowly increasing with $\xi/r_0$ among the $\tilde{b}$s given by Eqs. (\ref{tildeb1})-(\ref{tbquad}). 
In contrast, for $\alpha=5/3$ and, especially, for $\alpha=1$, $\tilde{b}$ exhibits a power-law increase with $\xi/r_0$, by Eqs. (\ref{bKappr}, \ref{tildeb1}), and attains finite values in the interval $\xi/r_0\lesssim 0.3$. The behaviour of the universal functions $\tilde{b}$ at leading order in $\xi/r_0$ is summarized in Table \ref{tab:alpha} below, for the three models of turbulence considered here.
\begin{table}%[H]
\centering
\renewcommand{\arraystretch}{2}
\caption{Relative crosstalk amplitude $\tilde{b}$ in the limit $\xi/r_0\ll 1$, as a function of $\xi/r_0$, for the three models of turbulence here considered.}
\begin{tabular}{c|c|c}
\vspace{-0.3cm}
	$\alpha=1$& $\alpha=5/3$ & $\alpha=2$\\
	 (linear)& (Kolmogorov)& (quadratic)\\
	\hline \hline
	$\tilde{b}\sim 1.20 \lt(\xi/r_0\rt)^{2}$& $\tilde{b}\sim 0.29 \lt(\xi/r_0\rt)^{8/3}$ & $\tilde{b}=\exp\lt(-\frac{\pi^2 r_0^2}{13.76\, \xi^2}\rt)$ \\
\end{tabular}
\label{tab:alpha} 
\end{table}

%\be
%\tilde{b}\approx 1.20 \lt(\xi/r_0\rt)^{2}\;(\alpha=1),\quad \tilde{b}\approx 0.29 \lt(\xi/r_0\rt)^{8/3}\; (\alpha=5/3).
%\e    

Finally, inserting (\ref{tildeb1}, \ref{tbK}, \ref{tbquad}) into (\ref{conc}), we obtain the universal entanglement decay laws for the three considered models of turbulence which we present in Fig.~\ref{fig3}. As anticipated, concurrence is most (least) robust against the rescaled turbulence strength for $\alpha=2$ ($\alpha=1$), with an intermediate behaviour for the Kolmogorov model. Apart from the three universal results here considered, we also plot the fitting function, \mbox{$g(x)=\exp(-4.16\,x^{3.24})$} which was obtained for $\alpha=5/3$ in \cite{leonhard15}. Although $g(x)$ is almost indistinguishable from the analytical result for finite values of $\xi/r_0$, our above discussion of the asymptotic behaviour of $\tilde{b}$ anticipates its limitations: Since $g(\xi/r_0)$ vanishes exponentially as $\xi/r_0\to \infty$,
Eq.~(\ref{conc}) implies $\tilde{b}(\infty)= 1/2$, in disagreement with the actual asymptotic limit $\tilde{b}(\infty)=1$. 

\section{Conclusion}
\label{sec:concl}
We studied the entanglement evolution of photonic orbital angular momentum (OAM)bipartite qubit states in a weakly turbulent atmosphere, for three models of turbulence which are distinguished by the exponent $\alpha=1,5/3,2$ of the phase structure function $D_\phi(x)=\gamma\,(x/r_0)^\alpha$. The entanglement evolution is entirely determined by the relative crosstalk amplitude $\tilde{b}$, such that an increase of $\tilde{b}$ entails a further loss of entanglement. $\tilde{b}$ is an $\alpha$-dependent, universal function of the ratio of the phase correlation length $\xi(l_0)$ (of an OAM beam with azimuthal index $l_0$ and radial index $p=0$) to the atmospheric transverse correlation length $r_0$. Using asymptotic methods, we obtained explicit analytical expressions for $\tilde{b}(\xi(l_0)/r_0)$. In particular, for small $\xi(l_0)/r_0\ll 1$, the relative crosstalk amplitude exhibits power-law dependencies $\sim (\xi(l_0)/r_0)^2$ and $\sim (\xi(l_0)/r_0)^{8/3}$ for $\alpha=1$ and $\alpha=5/3$, respectively, whereas it is exponentially suppressed for $\alpha=2$. 

Our results 
shed new light on some earlier, related work \cite{alpha13,Yan:16}.
In \cite{alpha13}, the concurrence decay of OAM bipartite qubit states \eqref{in_state} under Kolmogorov's turbulence ($\alpha=5/3$) was studied numerically and experimentally for odd $l_0\leq 7$. The observed behaviour was then compared to analytical results \cite{raymer06}, obtained for $\alpha=2$. Good agreement was found between numerical simulations and analytical predictions \cite{raymer06}. However, a faster entanglement decay under Kolmogorov's turbulence that is manifest in the simulations of \cite{alpha13} for $l_0>1$ -- in qualitative agreement with our present findings -- was not discussed in \cite{alpha13}. Although this feature was reported in numerical simulations \cite{Yan:16}, no analysis of the stronger robustness of entanglement for $\alpha=2$ was given. Curiously, although the experimental results of \cite{alpha13} exhibit significant scatter of the data points, for $l_0=5$ and $l_0=7$ most of the points lie below the predicted evolution of concurrence for $\alpha=2$. We conjecture that this trend will prevail if the experiment is carried out using states with larger azimuthal indices. Furthermore, it is noteworthy that there is a close correspondence between the results of \cite{alpha13,Yan:16}, where the output state was projected onto $p=0$, and of our present findings, where the output state was  averaged over the radial index $p$.  It therefore appears that the dependence of the concurrence decay on $\alpha$ dominates its dependence on the radial structure of the output state -- the latter being an open issue which should be addressed by future work.

Finally, our results suggest that the statistics of the wavefront distortions depends -- via $D_\phi(x)$ -- on the turbulence model. Therefore, another relevant future research topic is to identify the type of wavefront distortions resulting for distinct $\alpha$, especially for $\alpha=1$, wherein the entanglement decay is fastest. This information will potentially be useful when designing adaptive optics systems aimed at entanglement protection of twisted photons under weak to moderate turbulence 
\cite{PhysRevA.97.012321}.

%Besides it will help to resolve controversy between some studies suggesting that the largest entanglement loss is caused by the tilt of the wavefront \cite{Sheng:12} and simulations of the entanglement evolution in turbulence showing that the efficiency of adaptive optics in mending entanglement significantly enhances when one goes beyond the tip/tilt correction . 
\acknowledgements
Enjoyable and helpful discussions with Giacomo Sorelli are gratefully acknowledged. This work was supported by Deutsche Forschungsgemeinschaft under Grant DFG BU 1337/17-1.

\appendix
\section{Derivation of equation (\ref{defLambda})}
\label{Appmap}
Here, closely following \cite{NinaThesis}, we present a step by step derivation of the map operator (\ref{defLambda}) corresponding to a single phase screen. As mentioned above in section \ref{sec:model}, our model neglects scintillations and beam diffraction, and is thus limited to relatively short (of about $1$ km) propagation distances, for each photon. For convenience, we apply a single phase screen to a Laguere-Gaussian (LG) beam at $z=0$. Since the propagation through vacuum does not affect the beam's structure, we drop the $z$-dependence in all subsequent expressions, while including the propagation distance $L$ in the definition of the Fried parameter \eqref{eq:fried}.

Let us denote by $|p,l\rangle$ a single photon state populating the LG mode ${\rm LG}_{p,l}(r,\theta)$, i.e., 
\begin{equation} 
\langle r,\theta|p,l\rangle:={\rm LG}_{p,l}(r,\theta).
\label{LGsp}
\end{equation}
Any pure state $|\Psi_0\rangle$ can be expanded in terms of OAM basis states $|p,l\rangle$:
\begin{equation}
|\Psi_0\rangle=\sum_{p,l}c_{p,l}|p,l\rangle,
\label{eq:inexp}
\end{equation}
with $c_{p,l}$ given by the overlap integrals
\begin{equation}
c_{p,l}:=\frac{1}{2\pi}\int_{0}^{\infty}\int_{0}^{2\pi}{\rm LG}^*_{p,l}(r,\theta)\Psi_0(r,\theta)r \dif r \dif \theta,
\label{eq:apl}
\end{equation}
where $\Psi_0(r,\theta):=\langle r,\theta|\Psi_0\rangle$ is the mode function of the initial state.

The application of a phase screen to $|\Psi_0\rangle$ amounts to adding a random position-dependent phase $\phi(r,\theta)$ to the photon's wavefront:
\begin{equation}
|\Psi_0\rangle\rightarrow|\tilde{\Psi}\rangle=|\Psi_0\rangle e^{i\phi(r,\theta)}.
\label{eq:transf}
\end{equation}
Therefore, the expansion coefficients of $|\tilde{\Psi}\rangle$ in the OAM basis read
\begin{equation}
\tilde{c}_{p,l}:=\frac{1}{2\pi}\int_{0}^{\infty}\int_{0}^{2\pi}{\rm LG}^*_{p,l}(r,\theta)e^{i\phi(r,\theta)}\Psi_0(r,\theta)r \dif r \dif \theta.
\label{eq:bpl}
\end{equation}
Due to the phase factor $e^{i\phi(r,\theta)}$, the output state $|\tilde{\Psi}\rangle$ has an overlap with initially unpopulated OAM modes. In other words, phase aberrations lead to crosstalk among different OAM modes. 

Performing an ensemble average over phase screens with a given Fried parameter $r_0$, we obtain a mixed output state, 
\begin{align}
\varrho&:=\big\langle|\tilde{\Psi}\rangle\langle \tilde{\Psi}|\big\rangle=\sum_{p_1,l_1,p_2,l_2}\langle\tilde{c}_{p_1,l_1}\tilde{c}_{p_2,l_2}^*\rangle|p_1,l_1\rangle\langle p_2,l_2|\n\\
&=\sum_{p_1,l_1,p_2,l_2}\varrho_{p_1,l_1,p_2,l_2}|p_1,l_1\rangle\langle p_2,l_2|,
\label{eq:findens1}
\end{align}
with the matrix elements of the output state
\begin{align}
\varrho_{p_1,l_1,p_2,l_2}&=\frac{1}{4\pi^2}\int_{0}^{\infty}\hspace{-8pt}\int_{0}^{2\pi}\hspace{-8pt}\int_{0}^{\infty}\hspace{-8pt}\int_{0}^{2\pi}
r \dif r \dif \theta\,\, r' \dif r' \dif \theta'\n\\
&\times{\rm LG}^*_{p_1,l_1}(r,\theta){\rm LG}_{p_2,l_2}(r',\theta')\Psi_0(r,\theta)\Psi_0^*(r',\theta')\n\\
&\times \Big \langle \exp[i\phi(r,\theta)-i\phi(r',\theta')]\Big \rangle .
\label{eq:finden2}
\end{align}
Within the Kolmogorov turbulence model, the ensemble average in \eqref{eq:finden2} yields the result \cite{paterson05}
\begin{equation}
\Big \langle \exp[i\phi(r,\theta)-i\phi(r',\theta')]\Big \rangle=\exp\Big[-\frac{1}{2}D_\phi(|\vec{r}-\pvec{r}'|)\Big],
\label{dphi}
\end{equation}
where $D_\phi(|\vec{r}-\pvec{r}'|)$ is the {\it phase structure function} given by \cite{FRIED:65},
\begin{equation}
D_\phi(|\vec{r}-\pvec{r}'|)=6.88\left(\frac{|\vec{r}-\pvec{r}'|}{r_0}\right)^{5/3}.
\label{eq:dphi}
\end{equation}
We note that \eqref{eq:dphi} coincides with \eqref{eq:pls} for $\alpha=5/3$ and $x=|\vec{r}-\pvec{r}'|$.

From equations (\ref{LGsp}), (\ref{eq:inexp}), (\ref{eq:findens1})-(\ref{dphi}), the effect of the ensemble-averaged phase aberrations on any pure initial OAM state can be represented by the \emph{map operator} $\Lambda$ as follows:
\begin{equation}
\varrho_{p_1,l_1,p_2,l_2}=\sum_{p_0,l_0,p_0',l_0'}\Lambda_{p_1,l_1,p_2,l_2}^{p_0,l_0,p_0',l_0'}\,c_{p_0,l_0}c^*_{p_0',l_0'},
\label{eq:mapoperator0}
\end{equation}
where
\begin{align}
\Lambda_{p_1,l_1,p_2,l_2}^{p_0,l_0,p_0',l_0'}&=\frac{1}{4\pi^2}\int_{0}^{\infty}\hspace{-8pt}\int_{0}^{2\pi}\hspace{-8pt}\int_{0}^{\infty}\hspace{-8pt}\int_{0}^{2\pi}
r \dif r \dif \theta\,\, r' \dif r' \dif \theta'\n\\
&\times{\rm LG}^*_{p_1,l_1}(r,\theta) {\rm LG}_{p_2,l_2}(r',\theta') \n\\
&\times{\rm LG}_{p_0,l_0}(r,\theta) {\rm LG}_{p_0',l_0'}^*(r',\theta')\n\\
&\times \exp\left[-\frac{1}{2}D_\phi\left(|\vec{r}-\pvec{r}'|\right)\right].
\label{eq:mapoperator}
\end{align}
By factorizing the mode functions ${\rm LG}_{p,l}(r,\theta)=R_{p,l}(r)e^{il\theta}$
into products of their radial part $R_{p,l}(r)$ and a complex phase $e^{il\theta}$ \cite{siegman_lasers}, we obtain
\begin{widetext}
\begin{align}
\Lambda_{p_1,l_1,p_2,l_2}^{p_0,l_0,p_0',l_0'}&=\frac{1}{4\pi^2}\int_{0}^{\infty}\hspace{-8pt}\int_{0}^{2\pi}\hspace{-8pt}\int_{0}^{\infty}\hspace{-8pt}\int_{0}^{2\pi}R^*_{p_1,l_1}(r) R_{p_2,l_2}(r') R_{p_0,l_0}(r) R_{p_0',l_0'}^*(r')\n\\
&\times e^{-il_1\theta+il_2\theta'+il_0\theta-il_0'\theta'}\exp\left[-\frac{1}{2}D_\phi\left(|\vec{r}-\pvec{r}'|\right)\right] r \dif r \dif \theta\,\, r' \dif r' \dif \theta'.
\label{eq:mapoperator1}
\end{align}
\end{widetext}
To simplify this expression, we introduce new angular variables $\vartheta$ and $\Theta$,
\begin{equation}
\vartheta:=\theta-\theta',\quad \Theta:=\frac{\theta+\theta'}{2},
\label{angles}
\end{equation}
and take the integral over the angle $\Theta$, 
\begin{equation}
\int_{0}^{2\pi}e^{i(l_0-l_0'-l_1+l_2)\vartheta}\dif \Theta=2\pi\delta_{l_0-l_0',l_1-l_2}.
\end{equation}
Next, when tracing the output state over the radial quantum number $p$ (see section \ref{sec:model}),
\begin{equation}
\varrho_{l_1,l_2}=\sum_{p=0}^{\infty}\varrho_{p,l_1,p,l_2},
\label{eq:traceout}
\end{equation}
we focus on the special case $|l_1|=|l_2|=:l$. Therein, the symmetry holds,
\begin{equation}
R_{p,l}(r)=R_{p,-l}(r), 
\end{equation}
which follows from the general expression for LG functions \cite{siegman_lasers}. 
Finally, using the commutativity of the summation over $p$ and the integration over $r'$, we can employ the completeness relation \cite{abramowitz1966handbook}:
\begin{equation}
\sum_{p=0}^{\infty}R_{p,l}(r)R_{p,l}^*(r')=\frac{1}{r}\delta(r-r'),
\label{sump}
\end{equation}
to perform the integration over $r'$ in \eqref{eq:mapoperator1} and to arrive at the following expression for the matrix elements of the turbulence map: 
\begin{widetext}
\begin{equation}
\Lambda_{l,\pm l}^{p_0,l_0,p_0',l_0'}=\frac{1}{2\pi}\delta_{l_0-l_0',l\pm l}\int_{0}^{\infty}\hspace{-8pt}\int_{0}^{2\pi} R_{p_0,l_0}(r) R_{p_0',l_0'}^*(r)
 e^{-i\frac{\vartheta}{2}[(l\pm l)-(l_0+l_0')]}\exp\left\{-\frac{1}{2}D_\phi\left[2r\sin\left(\frac{\vartheta}{2}\right)\right]\right\} r \dif r \, \dif \vartheta,
\label{eq:mapoperatorl}
\end{equation}
\end{widetext}
which coincides with (\ref{defLambda}) for $p_0=p_0'=0$.
As we take the above steps, the argument $|\vec{r}-\pvec{r}'|$ of the phase structure function $D_\phi(|\vec{r}-\pvec{r}'|)$ undergoes the following transformations:
\begin{align}
|\vec{r}-\pvec{r}'|&=\sqrt{r^2+r'^2-2rr'\cos(\theta-\theta')}\\
&\stackrel{\eqref{angles}}{=}\sqrt{r^2+r'^2-2rr'\cos(\vartheta)}\n\\
&\stackrel{\eqref{eq:traceout}}{=}\sqrt{2r^2[1-\cos(\vartheta)]}\n\\
&=2r|\sin(\vartheta/2)|=2r\sin(\vartheta/2),
\end{align}
where the last equality follows, since $\sin(x)\geq0$ if $0\leq x\leq\pi$.

\section{Asymptotic evaluation of equation \eqref{defLambda}}
\label{appA}
%As shown in [], the cross-talk and survival amplitudes are expressed through the following matrix elements of the map $\Lambda$. 
Equation (\ref{defLambda}) can be rewritten in the form: 
\begin{align}
\Lambda_{l_1,l_1}^{l_0,l_0}&=\frac{2^{l_0+1}}{\pi l_0!}\int_{0}^{\infty}\hspace{-5pt}\int_{0}^{2\pi}\dif \rho \dif\vartheta e^{i\vartheta(l_1-l_0)}\n\\
&\times\exp\left[(2l_0+1)\ln(\rho)-2\rho^2-d(\vartheta)\rho^\alpha \right],
\label{eq:general}
\end{align}
where $\rho\equiv r/w_0$, $1\leq\alpha\leq 2$, and
\begin{equation}
d(\vartheta)=2^{\alpha-1}\gamma \left(\frac{w_0}{r_0}\right)^\alpha\sin^\alpha\left(\frac{\vartheta}{2}\right). 
\end{equation}
We first note that, for arbitrary $l_0$, $l_1$, $\alpha$, $w_0/r_0$, and for any fixed $\rho\geq 0$, the integral over $\vartheta$ in Eq. (\ref{eq:general}) is real. The corresponding integrand, $\exp\{-d(\vartheta)\rho^\alpha+i\vartheta(l_1-l_0)\}$, attains maximum values at the end points of the integration region and is symmetric  under reflection with respect to the $\vartheta=\pi$ axis. Therefore, when evaluating the integral over $\vartheta$, we can consider only the values $\vartheta\ll 1$. 
We use this consideration in our subsequent analysis of the integral over $\rho$ in Eq. (\ref{eq:general}), which is a Laplace integral \cite{erdelyi},
\begin{equation}
\int_0^\infty \dif\rho \exp[\lambda f(\rho)],
\label{lap_int}
\end{equation}
with 
\begin{equation}
f(\rho)=\ln\rho-\frac{2\rho^2-d(\vartheta)\rho^\alpha}{2l_0+1},
\label{frho}
\end{equation}
and $\lambda=2l_0+1$. For $l_0\gg1$, the asymptotics of the integral (\ref{lap_int}) can be evaluated using the method of steepest descent \cite{carrier2005functions,erdelyi}, according to which the main contribution to (\ref{lap_int}) comes from the neighborhood of the saddle points of the exponential $f(\rho)$.
The function $f(\rho)$ has a non-degenerate saddle point at
 \begin{equation}
 \rho_m\approx\sqrt{l_0/2},
\label{r_m}
 \end{equation}
where we neglected the term proportional to $d(\vartheta)$ in \eqref{frho} above. Inserting the expansion $f(\rho)\approx f(\rho_m)+f^{\prime\prime}(\rho_m)(\rho-\rho_m)^2$ into (\ref{eq:general}), extending the lower limit of the integral over $\rho$ to $-\infty$, and evaluating the resulting Gaussian integral, we arrive at the intermediate expression
\begin{equation}
\Lambda_{l_1,l_1}^{l_0,l_0}\approx\frac{1}{\pi}{\rm Re}\lt[\int_0^\pi \dif\vartheta \exp\{-A \vartheta^\alpha+i\vartheta(l_1-l_0)\}\rt],
\label{interm1}
\end{equation}
where 
\begin{equation}
A=\gamma 2^{-\alpha/2-1}\,l_0^{\alpha/2}\,(w_0/r_0)^\alpha.
\label{defA}
\end{equation}
%In writing Eq. (\ref{interm1}), we took into account the periodicity of the sine function and changed the upper integration limit from $2\pi$ to $\pi$, multiplying the whole expression by a factor of 2. Furthermore, we used an approximation
%\mbox{$d(\vartheta)\approx\gamma/2\,(w_0\vartheta/r_0)^\alpha$.} The periodicity of the integrand is destroyed upon this approximation, rendering the original real expression complex (for $l_1\neq l_0$). This violation is compensated by taking the real part of (\ref{interm1}).

For $l_1=l_0$ and $l_1=-l_0$,  Eq. (\ref{interm1}) yields the amplitudes $a$ and $b$, respectively. For the survival amplitude $a$, one immediately obtains,
\begin{equation}
a\approx\frac{1}{\pi}A^{-1/\alpha}\Gamma\lt(1+\frac{1}{\alpha}\rt).
\label{amplAa}
\end{equation}  
%or, using (\ref{defA}),
%\be
%a=\frac{1}{\pi}2^{1/2+1/\alpha}\gamma^{1/\alpha}l_0^{-1/2}\frac{r_0}{w_0}\Gamma\lt(1+\frac{1}{\alpha}\rt).
%\e
Since the main contribution to the crosstalk amplitude,
\begin{equation}
b\approx\frac{1}{\pi}{\rm Re}\lt[\int_0^\pi \dif\vartheta\exp(-A\vartheta^\alpha-2il_0\vartheta)\rt],
\label{defb}
\end{equation}
originates from $\vartheta\ll 1$, we can replace the upper integration limit by infinity, to obtain
\begin{equation}
b\approx\frac{1}{\pi}{\rm Re}\lt[\int_0^\infty \dif\vartheta\exp(-A\vartheta^\alpha-2il_0\vartheta) \rt]=:\frac{1}{\pi}{\rm Re}[J].
\label{defb1}
\end{equation}
%In this work, we are dealing with three particular values of $\alpha$: 1, 5/3, and 2. 
For $\alpha=1$ and $\alpha=2$, the oscillatory integral $J$ in Eq. \eqref{defb1} reduces to the Fourier transform of, respectively, an exponential and a Gaussian function. The parameter $b$ can be then easily calculated to yield Eqs. (\ref{ablin}) and (\ref{as_a_b_q}) for $\alpha=1$ and $\alpha=2$, respectively. For  $\alpha=5/3$, Eq. (\ref{defb1}) can be transformed to a tabulated integral, by a rotation of the real semi-axis by an angle $\beta$ into the complex plane \cite{carrier2005functions}. 
%To that end, Laplace-transform type integral using standard methods of complex analysis \cite{carrier2005functions}. The original integral over the real semi-axis is thereby converted to an integral in the complex plane along the line forming an angle $\beta$ with the real positive semi-axis. 
Then the integral $J$ transforms into
\begin{equation}
J=e^{i\beta}\int_0^\infty \dif x \exp(-q x^\alpha-sx),
\label{J_lap}
\end{equation} 
where $q=A\exp(i\alpha\beta)$ and $s=2il_0\exp(i\beta)$. The integral on the right hand side of Eq. (\ref{J_lap}) converges if ${\rm Re}(q)>0$, ${\rm Re}(s)>0$. To ensure that both inequalities are satisfied for $\alpha=5/3$ and for convenience, we choose $\beta=-\pi/5$. The exact value of Eq. (\ref{J_lap}) can be found in \cite{Prudnikov:1992:IS4}; it is expressed in terms of the Meijer G-function \cite{bateman}, see Eq. (\ref{bK}). We mention that the same method can also be used to find $b$ for other $\alpha=l/k$, where $l,k$ are positive integers, and in particular for $\alpha=1$ and $\alpha=2$. In the latter case, the Meijer G-function simplifies to the expressions for $b$ given by Eqs. (\ref{ablin}) and (\ref{as_a_b_q}), respectively. 

The relative crosstalk amplitude, derived directly from Eqs. (\ref{aK}) and (\ref{bK}),  reads 
\begin{align}
\tilde{b}&=\!\sqrt{\frac{3}{5}}\frac{3^{\frac{3}{5}}}{(2\pi)^3}\frac{1}{\Gamma\lt(\frac{8}{5}\rt)}\n\\
&\re\!\!\lt[z^{\frac{1}{5}}
G^{3,5}_{5,3}\lt(\!\Biggl.z\Biggr|\begin{array}{lllll}
0,&\!\frac{1}{5},&\!\frac{2}{5},&\!\frac{3}{5},&\!\frac{4}{5}\\
0,&\!\frac{1}{3},&\!\frac{2}{3}&&
\end{array}
\rt)\!
\rt],
\label{tildebK}
\end{align}
where 
\begin{align*}
z=-i\frac{5^5t^5\gamma^3}{3^32^{21/2}l_0^{5/2}}.
\end{align*}

%Like in the case $\alpha=1$, we compare the exact numerical and asymptotic expressions for $\alpha=5/3$ in Fig. +(b). 
 
To show that $\tilde{b}$ is a function of the sole parameter $\xi/r_0$, we use the identity \cite{bateman}
\begin{equation}
z^\sigma G^{m,n}_{p,q}\left(\Biggl.z\Biggr|\begin{array}{l}
a_r\\
b_{r^\prime}
\end{array}
\right)=G^{m,n}_{p,q}\lt(\Biggl.z\Biggr|\begin{array}{l}
a_r+\sigma\\
b_{r^\prime}+\sigma
\end{array}
\right),
\label{bate_id}
\end{equation}
where $a_r$ ($b_{r^\prime}$) represent the upper (lower) vectors of the Meijer G-function (e.g., in Eq. (\ref{tildebK}), \mbox{$a_r=\{0,1/5,2/5,3/5,4/5\}$)} and the summation in the right-hand side of \eqref{bate_id} is to be understood element-wise. Applying the identity (\ref{bate_id}) to Eq. (\ref{tildebK}), we arrive at the 
result
\begin{align}
\tilde{b}&=\sqrt{\frac{3}{5}}\frac{3^{3/5}}{(2\pi)^3}\frac{1}{\Gamma\lt(\frac{8}{5}\rt)}\label{tbKapp}\\
&\times\!\!\re\!\!\lt[
G^{3,5}_{5,3}\lt(\Biggl.\frac{-i\gamma^35^5\xi^5}{\pi^56^3r_0^5}\Biggr|\begin{array}{lllll}
\frac{1}{5},&\frac{2}{5},&\frac{3}{5},&\frac{4}{5},&1\\
\frac{1}{5},&\frac{8}{15},&\frac{13}{15}&&
\end{array}
\rt)
\rt].\n
\end{align}

Whereas the limiting values $0$ and $1$ of $\tilde{b}$ for $\xi\to 0$ and $\xi\to \infty$, respectively, can be directly computed from Eq. (\ref{tbKapp}), the asymptotics of $\tilde{b}$ for $\xi/r_0\ll 1$ (or \mbox{$w_0/r_0\sqrt{l_0}\ll 1$)} is easier to deduce directly from Eqs. (\ref{amplAa}), (\ref{defb1}) and (\ref{J_lap}). It follows from (\ref{amplAa}) and (\ref{defb1}) that $\tilde{b}=\re[J]/(\pi a)$. Using Watson's lemma \cite{carrier2005functions}, we obtain the asymptotic series expansion of the integral (\ref{J_lap}):
\begin{equation}
J=e^{i\beta}\frac{1}{s}\sum_{n=0}^\infty \frac{(-1)^n}{n!}\lt(\frac{q}{s^\alpha}\rt)^n\Gamma(1+n\alpha).
\label{Jseries}
\end{equation}  
Substituting the values $\alpha=5/3$, $\beta=-\pi/5$, $s=2il_0\,e^{-i\pi/5}$, $q=6.88\times2^{-11/6}\,t^{5/3}\,l_0^{5/6} e^{-i\pi/3}$ and $a=\pi^{-1}q^{-3/5}e^{-i\pi/5}\Gamma(1+3/5)$ into the ratio $\tilde{b}=\re[J]/(\pi a)$, after some simple algebra we obtain 
\begin{align}
\tilde{b}&\approx 0.380\!\lt(\!\frac{t^2}{l_0}\!\rt)^{\!\!\frac{4}{3}}\!\!\!\!+\!1.231\!\lt(\!\frac{t^2}{l_0}\!\rt)^{\!\!\frac{13}{6}}\!\!\!\!\!\!+\!3.735\!\lt(\!\frac{t^2}{l_0}\!\rt)^{\!\!3}\!\!\!+\!O\!\!\lt[\lt(\!\frac{t^2}{l_0}\!\rt)^{\!\frac{23}{6}}\rt]\n\\
&=0.287 \!\lt(\!\frac{\xi}{r_0}\!\rt)^{\!\!\frac{8}{3}}\!\!\!+\!0.781\!\lt(\!\frac{\xi}{r_0}\!\rt)^{\!\!\frac{13}{3}}\!\!\!\!\!\!+\!1.989\!\lt(\!\frac{\xi}{r_0}\!\rt)^{\!\!6}\!\!\!+\!O\!\lt[\lt(\!\frac{\xi}{r_0}\!\rt)^{\!\frac{23}{3}}\rt],
\label{As_ser}
\end{align}
where the top line reproduces \eqref{As_ser1} and the bottom line is used in Fig. \ref{fig2}. Note an increase of the subsequent expansion coefficients in \eqref{As_ser}. Starting from some $n$, which value depends on $t$ and $l_0$, these coefficients dominate the decreasing magnitude of $(t^2/l_0)^{(4+5n)/3}$ in the asymptotic expansion of $\tilde{b}$ (for $t=5.0$ and $l_0=150$, this occurs at $n=7$). This behaviour indicates divergence of Eq. (\ref{Jseries})  -- the common feature of asymptotic series \cite{dingle}. However, a finite number of terms in Eq. (\ref{Jseries}) yields an accurate asymptotic description of $\tilde{b}$. We empirically established that the first three terms of the series \eqref{As_ser} provide the best overlap with the exact behaviour of $\tilde{b}$ at $l_0\gtrsim 150$ and $0.1 \leq t\leq 5.0$ (i.e., for $0.0091\leq \xi/r_0\leq 0.45$), see Figs.\ref{fig1} and \ref{fig2}.

\bibliography{OAM_db1}
\bibliographystyle{apsrev4-1}

\end{document}